\documentclass[traditabstract]{aa}

\usepackage{amsmath}
\usepackage{natbib}
\usepackage{graphicx}
\usepackage{microtype}
\usepackage{url}
\usepackage{txfonts}
\usepackage{color}
\usepackage{xspace}

\newcommand{\keV}{\ensuremath{\text{ke\kern -0.09em V}}\xspace}
\newcommand{\grs}{GRS\,1758$-$258\xspace}
\newcommand{\cyg}{Cyg~X-1\xspace}
\newcommand{\einsE}{1E~1740.7$-$2942\xspace}
\newcommand{\xte}{\textsl{RXTE}\xspace}
\newcommand{\chisq}{\ensuremath{\chi^2_{\mathrm{red}}}}

\newcommand{\nH}{\ensuremath{N_{\mathrm{H}}}}

\authorrunning{Hirsch et al.}
\titlerunning{Variability of GRS\,1758$-$258}

\title{X-ray spectral and flux variability of the microquasar \grs on
  timescales from weeks to years} 

\author{
 Maria Hirsch \inst{\ref{affil:remeis}}
 \and Katja Pottschmidt \inst{\ref{affil:cresst},\ref{affil:umbc}}
 \and David M.\ Smith \inst{\ref{affil:ucsc}}
 \and \mbox{Arash Bodaghee}\inst{\ref{affil:georgia}}
 \and Marion Cadolle Bel \inst{\ref{affil:marion}}
 \and Victoria Grinberg \inst{\ref{affil:tuebingen}}
 \and Natalie Hell \inst{\ref{affil:llnl}}
 \and Felicia Krau\ss{} \inst{\ref{affil:ams},\ref{affil:pen}}
 \and Ingo Kreykenbohm \inst{\ref{affil:remeis}}
 \and Anne Lohfink \inst{\ref{affil:montana}}
 \and Michael A.\ Nowak \inst{\ref{affil:mike}}
 \and \mbox{B\'arbara H.\ Rodrigues} \inst{\ref{affil:brasilcfa}}
 \and Roberto Soria\inst{\ref{affil:china},\ref{affil:sydney}}
 \and John A.\ Tomsick \inst{\ref{affil:ssl}}
 \and J\"orn Wilms \inst{\ref{affil:remeis}}
}

\institute{
      Dr.\ Karl Remeis-Observatory \& ECAP, Universit\"at 
      Erlangen-N\"urnberg, Sternwartstr.~7, 96049 Bamberg, Germany
      \label{affil:remeis}
 \and CRESST and NASA Goddard Space Flight Center, Greenbelt, MD 20771, USA
      \label{affil:cresst}
 \and Department of Physics and Center for Space Science and Technology, 
      University of Maryland, Baltimore County, Baltimore, MD 21250, USA
      \label{affil:umbc}
 \and Department of Physics and Santa Cruz Institute for Particle Physics,
      University of California, Santa Cruz, CA 95064, USA
      \label{affil:ucsc}
 \and Department of Chemistry, Physics and Astronomy, 
      Georgia College and State University, Milledgeville, GA 31061, USA
      \label{affil:georgia}
 \and Max Planck Computing and Data Facility, Gie\ss{}enbachstr.~2, 85748
      Garching, Germany
      \label{affil:marion}
 \and Institut f\"ur Astronomie und Astrophysik, Universit\"at T\"ubingen, 
      Sand 1, 72076 T\"ubingen, Germany
      \label{affil:tuebingen}
 \and Lawrence Livermore National Laboratory, 7000 East Ave., 
      Livermore, CA 94550, USA 
      \label{affil:llnl}
 \and GRAPPA \& Anton Pannekoek Institute for Astronomy, University of
      Amsterdam, Science Park 904, 1098 XH Amsterdam, The Netherlands
      \label{affil:ams}
 \and Department of Astronomy and Astrophysics, Pennsylvania State
      University, University Park, PA 16802, USA
      \label{affil:pen}
 \and Department of Physics, Montana State University, 
      Bozeman, MT 59717-3840, USA
      \label{affil:montana}
 \and Department of Physics, Washington University, CB 1058, 
      One Brookings Drive, St. Louis, MO 63130-4899, USA
      \label{affil:mike}
 \and Instituto Nacional de Pesquisas Espaciais, Avenida dos
      Astronautas, 1.758, S\~ao Jos\'e dos Campos, Brazil 
      \label{affil:brasilcfa}
 \and National Astronomical Observatories, Chinese Academy of Sciences, 
      Beijing 100012, China
      \label{affil:china}
 \and Sydney Institute for Astronomy, School of Physics A28, 
      The University of Sydney, Sydney, NSW 2006, Australia
      \label{affil:sydney}
 \and Space Sciences Laboratory, 7 Gauss Way,
      University of California, Berkeley, CA 94720-7450, USA
      \label{affil:ssl}
}

\date{Received  / Accepted}
\keywords{X-rays: individual \grs -- X-rays: binaries}

\begin{document}

\abstract{We present the spectral and timing evolution of the
  persistent black hole X-ray binary \grs based on almost 12 years of
  observations using the \textsl{Rossi X-ray Timing Explorer}
  Proportional Counter Array. While the source was predominantly found
  in the hard state during this time, it entered the thermally
  dominated soft state seven times. In the soft state \grs shows a
  strong decline in flux above 3\,keV rather than the pivoting flux
  around 10\,keV more commonly shown by black hole transients. In its
  3--20\,keV hardness intensity diagram, \grs shows a hysteresis of
  hard and soft state fluxes typical for transient sources in
  outburst. The \xte-PCA and \xte-ASM long-term light curves do not
  show any orbital modulations in the range of 2 to 30\,d. However, in
  the dynamic power spectra significant peaks drift between
  18.47 and 18.04\,d for the PCA data, while less significant
  signatures between 19\,d and 20\,d are seen for the ASM data as well as
  for the \textsl{Swift}/BAT data. We discuss different models for the
  hysteresis behavior during state transitions as well as
  possibilities for the origin of the long term variation in the
  context of a warped accretion disk.}

\maketitle

\section{Introduction}

\object{GRS\,1758$-$258} is a black hole binary discovered in 1990
during observations of the Galactic Center region by the
\textsl{Granat} satellite \citep[][see \citealt{heindl:02} for the
determination of the source's position]{mandrou:90,syunyaev:91}. As
one of only three persistent, mostly hard state, black hole binaries
in our Galaxy and the Magellanic Clouds (\grs, 1E\,1740.7$-$2942, and
\cyg)\footnote{Other persistent black hole binaries are the Galactic
  source 4U\,1957$+$11, as well as LMC~X-1 and LMC~X-3, which are
  always or predominantly found in the soft state.}, \grs has since
been observed in various energy ranges \citep[e.g.,][and references
therein]{rodriguez:92, cadolle:06, pottschmidt:08, munoz:10,
  soria:11,luque:14}. As radio observations show a double-lobed
counterpart \citep{rodriguez:92} that shows similarities to winged
radio galaxies \citep{marti:17}, \grs is considered a microquasar.

The \textsl{Rossi X-ray Timing Explorer} \citep[\xte;][]{bradt:93}
monitored \grs from 1996 to 2007 (see next section for a detailed
description of these observations). Based on this program
\citet{smith:01} reported the transition to a soft state in 2001
during which the 3--25\,keV flux declined by more than an order of
magnitude. This episode lasted around a year. Using
\textsl{XMM-Newton} observations \citet{soria:11} showed that this
transition still followed the canonical evolution through states but
with the soft state showing increased flux only below $\sim$3\,keV.
The occurrence of a less extended soft state in 2003 was reported by
\citet{pottschmidt:06}.

Using the 2.5--25\,keV \xte PCA monitoring light curve of \grs from
1997 to 2002, \citet{smith:02} found a periodic signal of
$18.45\pm 0.10$\,d. If this signal was due to a binary orbit and the
companion is not a high-mass star \citep{marti:98}, the companion
would have to be a K giant star filling its Roche lobe
\citep{rothstein:02}. However, the identity of the companion star in
the system was ambiguous for a long time \citep{smith:10}. After
astrometric studies already hinted at the system being an
intermediate-mass X-ray binary \citep{munoz:10,luque:14}, recent
spectroscopy of the companion shows it is likely an A-type main
sequence star \citep{marti:16}. If true, the accretion process would
still be Roche lobe overflow, which implies an orbital period in the
range of 0.5--1.0\,d \citep{marti:16}, and, thus, significantly
shorter than the signal at $18.45\pm0.10$\,d found by
\citet{smith:02}.

In order to better understand this puzzling source we took a closer
look at the \xte monitoring obervations of \grs for the first time
spanning the full time range from 1996 to 2007. The observations and
the Proportional Counter Array \citep[PCA;][]{jahoda:06} data
reduction are described in Sect.~\ref{sec:obs}. In
Section~\ref{sec:spectral}, we present the spectral analysis of the
\xte-PCA data of \grs, starting with a description of how Galactic
Ridge emission was taken into account in Sect.~\ref{sec:bg} before
defining the source model and overall spectral modeling procedure in
Sect.~\ref{sec:specmodel}, and adding the hardness intensity diagram
(HID) of the dataset as an additional diagnostic in
Sect.~\ref{sec:hid}. The evolution of the spectral parameters as well
as the HID show seven soft states during this time. In
Sect.~\ref{sec:timing}, we report on the timing analysis of the
long-term light curve of \grs, presenting periodograms of the raw as
well as of the detrended \xte-PCA light curves in
Sect.~\ref{sec:psdraw} and~\ref{sec:scargle}. Finally, we also looked
at periodograms of \xte's All Sky Monitor \citep[ASM;][]{levine:96}
light curve of \grs in Sect.~\ref{sec:psdasm}. We find no coherent
periodic signal that could be identified as an orbital period in the
range of 2\,d to 30\,d, however, quasi-periodic oscillations are
apparent in the period range around 18--20\,d. In
Sect.~\ref{sec:summary}, we summarize and discuss the results and
present our conclusions.

\section{Observations and data reduction}\label{sec:obs}

\xte monitored \grs with 1.0--1.5\,ks long pointed snapshots starting
in 1996 \citep{smith:01,smith:02}. The exposures were performed in
monthly intervals in 1996, weekly from 1997 through 2000, and twice a
week from 2001 March to 2007 October. Each year there is a gap from
November to January when the Sun was too close to the Galactic Center,
that is the approximate pointing direction to \grs.

The PCA consisted of five Proportional Counter Units (PCUs), each
sensitive between 2\,keV and 90\,keV with a field of view of
$\sim$$1^\circ$. Proportional Counter Unit 2 (PCU2) was the best
calibrated one \citep{jahoda:06}. Since the top layer had the highest
signal to noise ratio, we only used data from this layer of PCU2. We
reduced the data with our standard analysis pipelines \citep{wilms:99,
  wilms:06} applying the NASA HEASARC software package
\textsc{heasoft}, version 6.8 for the \xte spectra.\footnote{As of the
  time of writing, \textsc{heasoft} had not changed in parts relevant
  for \xte data reduction and calibration since this release. A
  difference of 0.4\% in total flux for PCU2 with respect to later
  \textsc{heasoft} versions is due to an improved cross-calibration of
  the PCUs. This change does not affect our analysis.} Data up to
15\,minutes after passage through the South Atlantic Anomaly (SAA)
were excluded \citep{fuerst:09}. We also required an electron ratio
below 0.5 in order to exclude time periods of high background. We
obtained observed spectra as well as instrumental background spectra
using the ``faint source'' background model.

Because of its location only $0\fdg66$ away from the very bright X-ray
binary GX\,5$-$1, \grs was a difficult source to observe with the
collimated detectors onboard \xte. The monitoring could therefore only
be realized using offset pointings away from GX\,5$-$1
\citep{smith:01,smith:02}, that is using the triangular response of
PCA's collimator to reduce the influence of GX\,5$-$1. Response
matrices were built taking the effect of the offset pointings into
account.

The spectral analysis was performed using data taken in standard2f
mode, which provides 129 energy channels. The spectra were rebinned to
a minimum signal to noise ratio of 5. We used data in the energy range
of 3--20\,keV. All spectral fitting was done using the
\textsl{Interactive Spectral Interpretation System}
\citep[ISIS;][]{houck:00, houck:02, noble:08}.

The PCA timing analysis was performed starting from the 3--20\,keV
fluxes of \grs determined for each monitoring observation in the
spectral analysis. The ASM timing analysis was performed using the
3--5\,keV light
curve\footnote{\url{ftp://legacy.gsfc.nasa.gov/xte/data/archive/ASMProducts/definitive\_1dwell/colors/xa\_grs1758-258\_d1.col}}
available in NASA's HEASARC database in a daily binning for the same
time range as the PCA light curve. Empty bins were removed. The
\textsl{Neil Gehrels Swift Observatory} BAT 15--50\,keV daily light
curve was obtained from the BAT Transient Monitor
\citep{krimm:13}\footnote{\url{https://swift.gsfc.nasa.gov/results/transients/GRS1758-258/}}.
Both the ASM and the BAT light curve do not require further processing
before the analysis. Due to the location of \grs near the GX\,5$-$1,
\textsl{MAXI} \citep{matsuoka09a} cannot observe the source.

\section{Spectral evolution}\label{sec:spectral}

\subsection{Galactic Ridge emission}\label{sec:bg}

\begin{figure}
  \resizebox{\hsize}{!}{\includegraphics{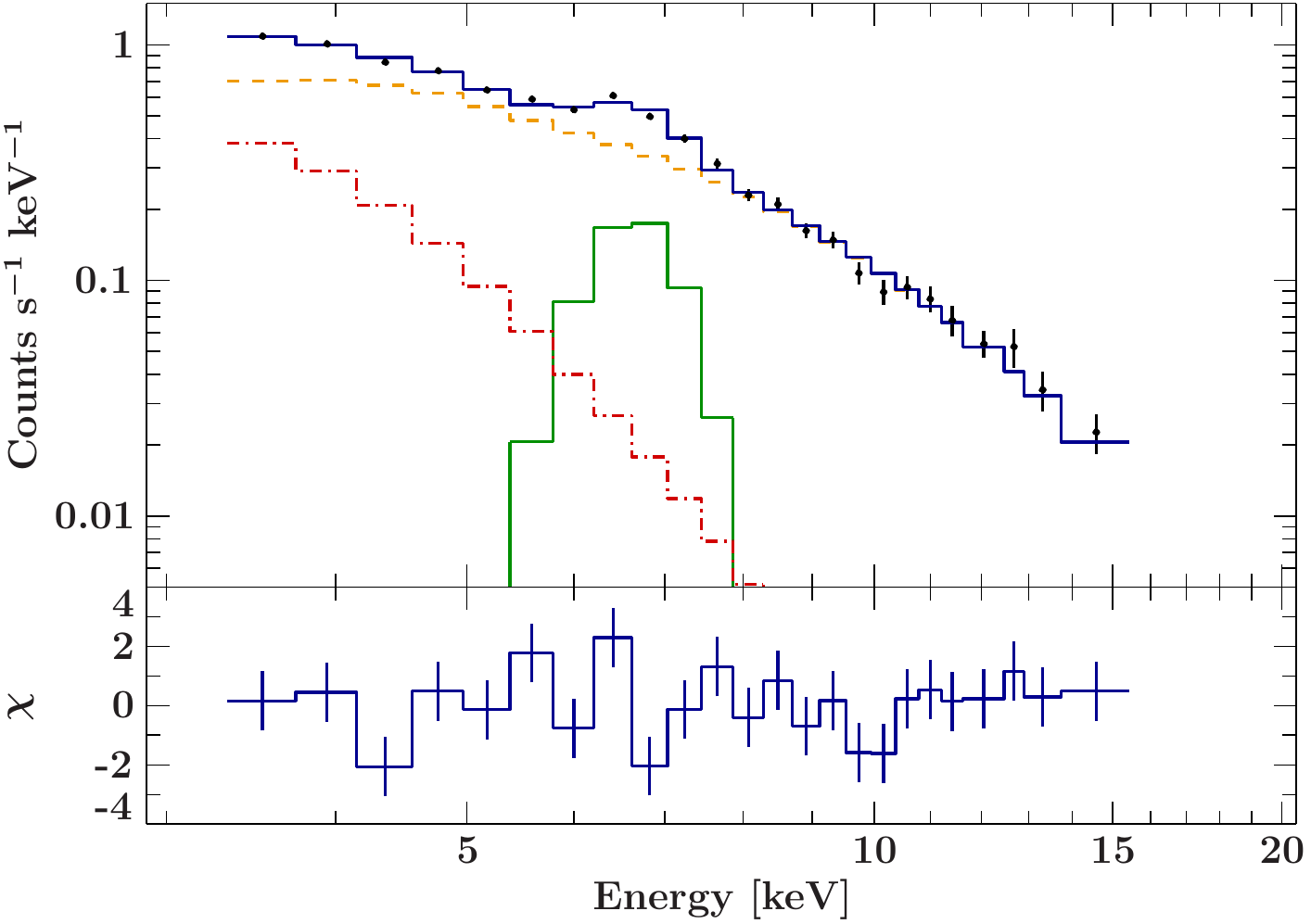}}
  \caption{Spectrum of the Galactic Ridge emission as seen by \xte.
    The data were fitted with two bremsstrahlung components (1: dashed
    line, 2: dash-dotted line) and an iron line complex as described
    in \citet{ebisawa:07}.}
  \label{fig:ridge_comp}
\end{figure}
\begin{figure}
\includegraphics[width=\columnwidth]{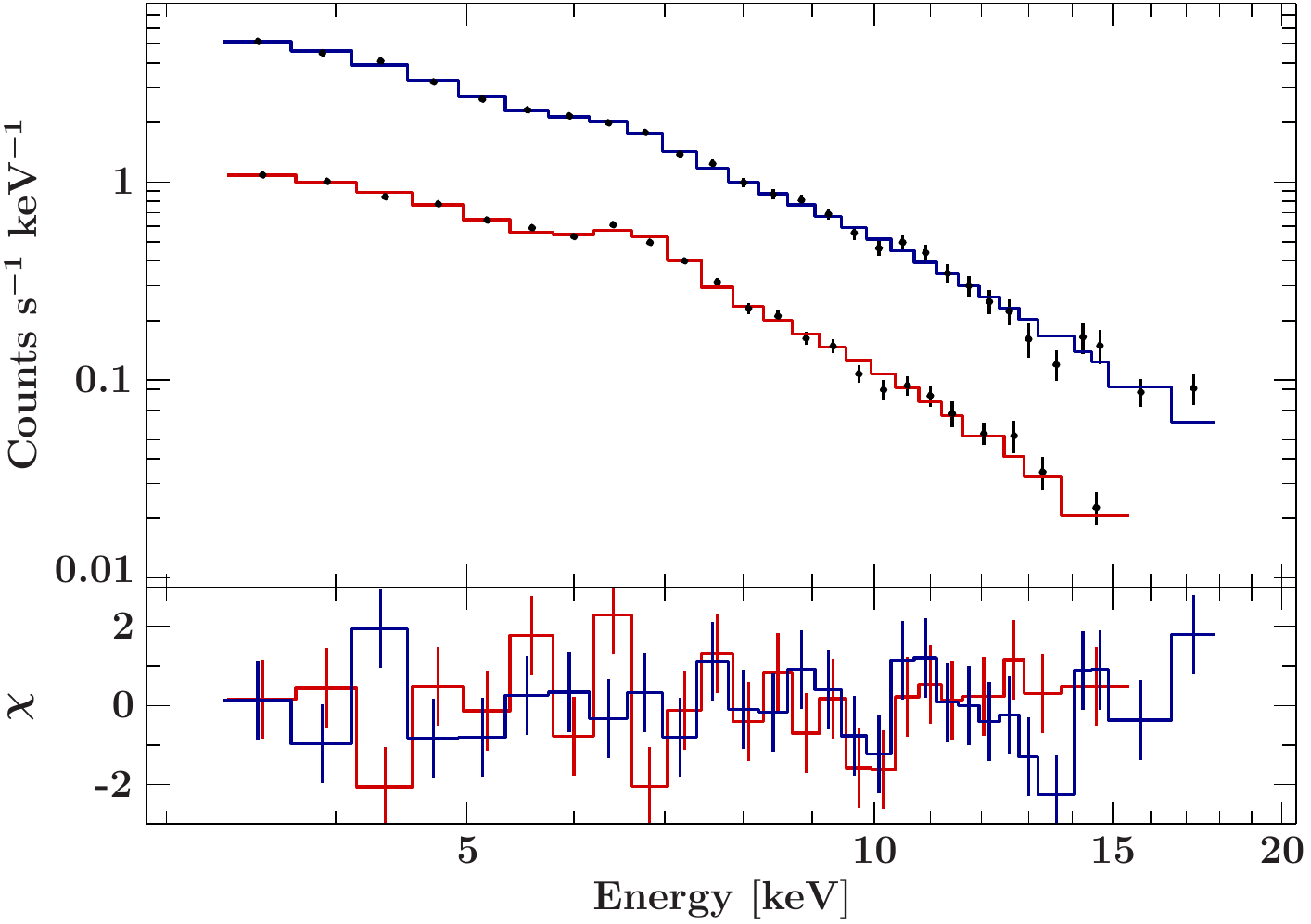}
\caption{\xte-PCA spectrum of the total emission at the source
  position from the 2003 April observation. The total spectrum is
  modeled (blue histogram) as the sum of the source contribution and
  the Galactic ridge emission. The contribution of the the Galactic
  ridge emission alone is also shown (red histogram).}
\label{fig:ridge_source}
\end{figure}

Because \grs is faint and located close to the Galactic center in the
Galactic Plane, all \xte spectra of the source also contain a strong,
diffuse background component caused by the Galactic Ridge Emission in
the X-rays. This emission has long been known to exist
\citep{worrall:82,warwick:85,koyama:86}, its origin, however, is still
under discussion \citep[][and references therein]{ebisawa:08,
  warwick:14, nobukawa:16}. To distinguish between source counts and
Galactic Ridge counts, background observations totalling 13\,ks,
$1\fdg5$ offset from \grs, were performed by \xte in
1999\footnote{ObsIDs \mbox{40097-09-01-00}, \mbox{40097-09-02-00},
  \mbox{40097-09-02-01}, \mbox{40097-09-02-02}, \mbox{40097-09-02-03},
  \mbox{40097-09-03-00}, and \mbox{40097-09-04-00}}. We were able to
model this local Galactic Ridge Emission with two bremsstrahlung
components and an iron line complex (see Fig.~\ref{fig:ridge_comp}).
As \xte cannot resolve the individual iron line components, the
position of the three lines was fixed to 6.4\,keV, 6.67\,keV, and
7.0\,keV, respectively, with equivalent widths that scale as
$85:458:129$ according to CCD \textsl{Suzaku} observations of the
Galactic Ridge \citep{ebisawa:07}. The normalization of the whole
complex was left free to vary in the fit. The fit parameters are
summarized in Table~\ref{tab:ridge}. We assume that there is no local
variation of the Galactic Ridge emission, and then, keeping all
spectral parameters fixed at their best-fit values, added this model
to the spectral model of \grs. Figure~\ref{fig:ridge_source}
illustrates the contribution of the ridge emission to the measured
spectrum.

\begin{table}
  \caption{Galactic Ridge model parameters.}
  \label{tab:ridge}
  \centering
  \renewcommand{\arraystretch}{1.3}%
  \begin{tabular}{lll}
    \hline
    \hline
    $a_{\mathrm{brems},1}$   & $0.011\pm0.003$            & \\
    $kT_{1}$             & $8^{+3}_{-1}$                & \keV\\
    $a_{\mathrm{brems},2}$   & $0.05^{+0.03}_{-0.01}$        & \\
    $kT_{2}$             & $1.2^{+0.2}_{-0.1}$            & \keV\\
    \hline
    $F_1$              & $\left(2.6^{+0.3}_{-0.4}\right)\times10^{-5}$   & $\mathrm{ph}\,\mathrm{s}^{-1}\,\mathrm{cm}^{-2}$\\ 
    $\sigma_{1}$                  & $0.05$                      & \keV\\
    $E_1$              & $6.4$                       &
    \keV\\
    $F_2$              & $1.4\times10^{-4}$           & $\mathrm{ph}\,\mathrm{s}^{-1}\,\mathrm{cm}^{-2}$\\
    $\sigma_2$                  & $0.05$                      & \keV\\
    $E_{2}$              & $6.67$                      & \keV\\
    $F_{3}$              & $4\times10^{-5}$             & $\mathrm{ph}\,\mathrm{s}^{-1}\,\mathrm{cm}^{-2}$\\
    $\sigma_3$                  & $0.05$                      & \keV\\
    $E_3$              & $7.0$                       & \keV\\
    \hline
  \end{tabular}
  \tablefoot{$a_{\mathrm{brems},1,2}$: normalization of the
    bremsstrahlung components; $kT_{1,2}$: plasma temperature;
    $F_{1,2,3}$ is the total flux under a Gaussian centered at
    $E_{1,2,3}$ with a width $\sigma_{1,2,3}$. Values without
    uncertainties were kept fixed during the fit to the spectra from
    the background regions. Uncertainties are at the 90\% confidence
    level.}
\end{table}

\subsection{Spectral modelling}\label{sec:specmodel}

Once the Galactic Ridge background has been accounted for (see
Sect.~\ref{sec:bg}), all spectra were modeled using an empirical model
consisting of an absorbed powerlaw (\texttt{phabs $\times$ powerlaw}).
No high energy cutoff was needed as the cutoff energy is well above
20\,keV \citep{pottschmidt:06}, the upper limit of the energy range
considered here. The column density due to interstellar absorption in
the direction of \grs was kept fixed to the canonical value of
$\nH = 1.5 \times 10^{22}\,\mathrm{cm}^{-2}$ \citep{mereghetti:97}
using the abundances of \citet{anders:89}. Although the Galactic ridge
spectral component already contains an iron line complex, the
residuals show that this component is insufficient to explain the data
in the Fe K$\alpha$ band. There are three potential explanations for
such a deviation: The Galactic ridge emission could be spatially
variable, there could be an intrinsic Fe K$\alpha$ emission, or a
combination of both effects.

In order to characterize the deviation of the observed line from the
assumed (non-spatially varying) ridge component, we add a line at a
fixed energy of 6.4\,keV to the model. We initially fixed the width of
this line at 1\,eV, well below the resolution of the PCA, and then
fitted the flux of the line. In these fits a possible weak correlation
between iron line flux and total source flux with a Spearman rank
coefficient of 0.40 can be seen. A rough check of 100000 permutations
of the iron line flux against the total source flux gives a mean rank
coefficient of $7.9 \times 10^{-5}$ with individual values ranging
between $-0.15$ and 0.15. If true, this variability would indicate
that part of the line would be source intrinsic. We note, however,
that a narrow line at 6.4\,keV with a flux similar to that found in
these fits would have been visible in the \textsl{XMM-Newton}
observations discussed by \citet{soria:11}, but was not seen.
\textsl{RXTE}'s spectral resolution is so low, however, the width of
the additional Fe K$\alpha$ is not well constrained: Re-fitting the
\xte spectra with the iron line width left as a free parameter results
in an average line width of about 800\,eV. Simulating
\textsl{XMM-Newton} EPIC~pn spectra for some of our best fit models
with such a broad line smears out the iron line beyond recognition.
For this reason we cannot claim that the \textsl{XMM-Newton} data
formally rule out that some of the broad line flux originates in \grs,
although we consider this an astrophysically unlikely interpretation
of the \xte result.

\begin{figure}
  \resizebox{\hsize}{!}{\includegraphics{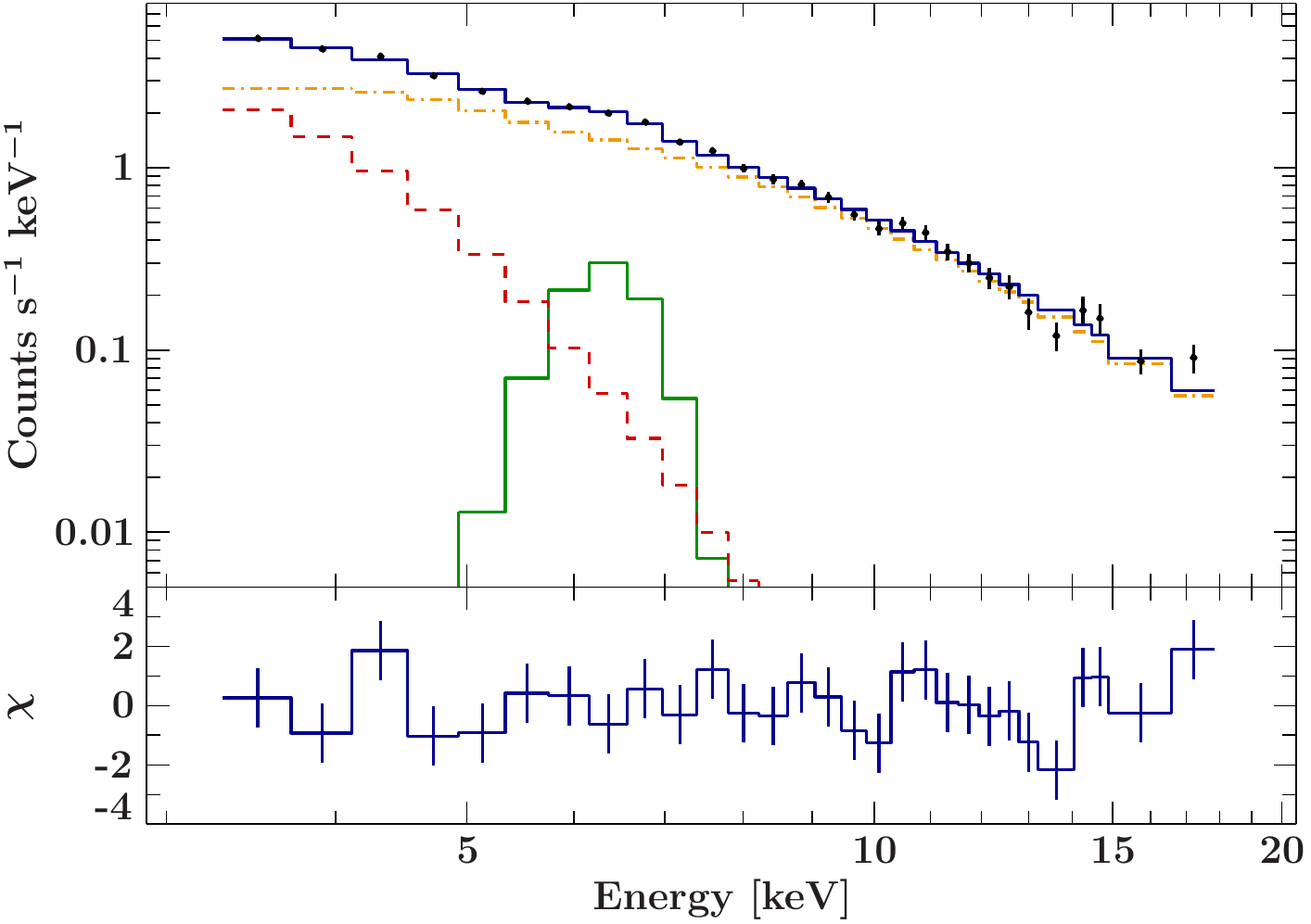}}
  \caption{Example of an instrument background subtracted spectrum
    taken by \xte on 2009 April 08, containing the absorbed powerlaw
    component (dash-dotted line), the disk (dashed line), and the iron
    line (green solid line). For clarity the constant Galactic Ridge
    model part is not shown.}
  \label{fig:data}
\end{figure}

Some softer spectra also require an additional disk blackbody for
improving the fit. To estimate the significance of this improvement,
we performed Monte Carlo simulations of the best fit model without the
disk by creating a set of 1000 fake spectra for each observation.
These synthetic spectra were then fitted with both models and the
respective improvement in \chisq was calculated. We only accepted the
disk component in our best fit model if the improvement in \chisq of
the real dataset was above at least 99\% of the fake spectra
improvements. There are four occasions where a disk is detected at a
very high temperature and low normalization. All four can be modeled
with a higher normalization and lower temperature, with only slight
worsening in the reduced $\chi^2$. These outliers thus are likely to
be reflecting fit degeneracies and are not considered to be source
intrinsic. Figure~\ref{fig:data} shows an example of an instrument
background subtracted spectrum and best fit model containing all the
components.

\begin{figure*}
  \centering
  \includegraphics[width=17cm]{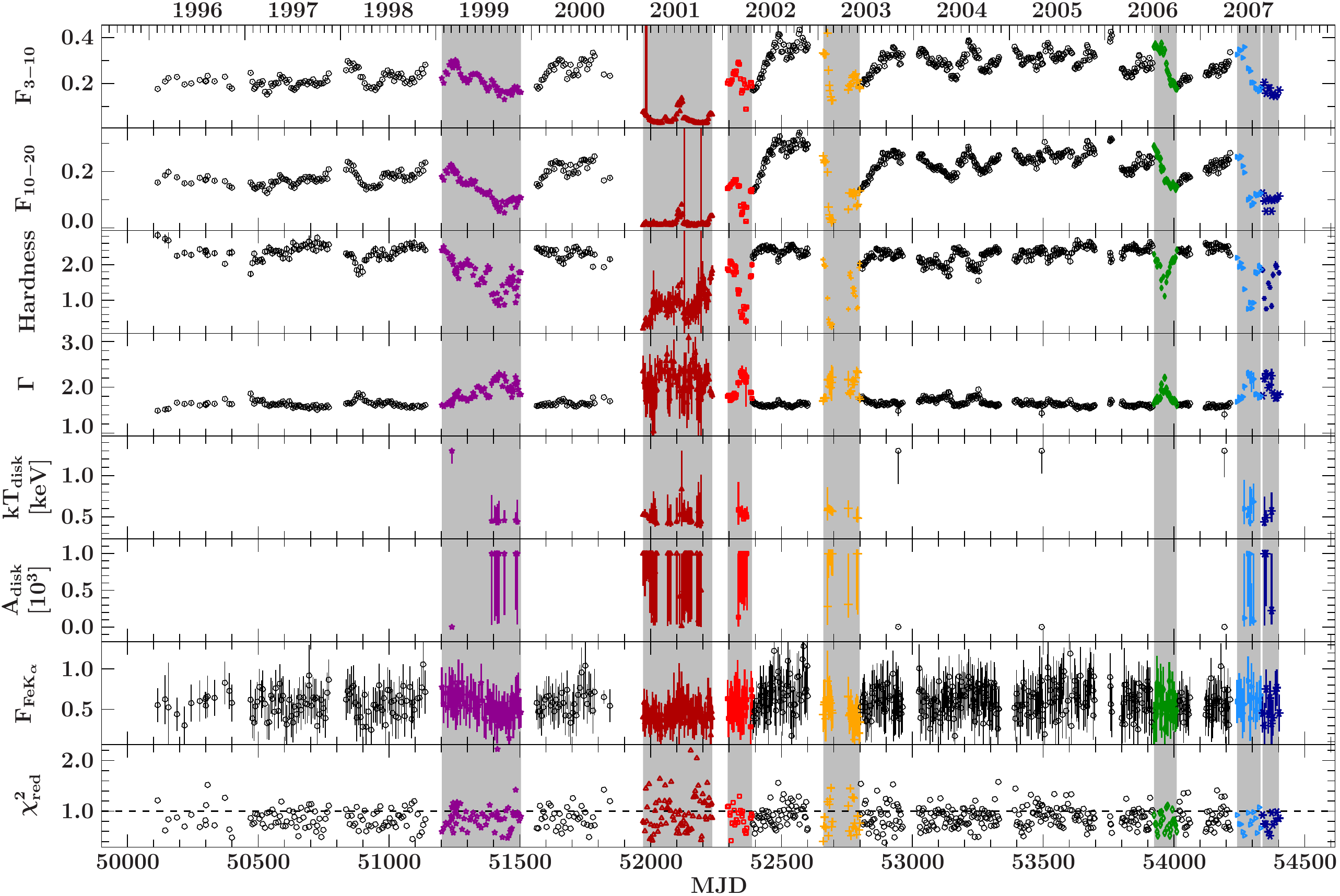}
  \caption{Spectral parameters from \xte monitoring observations of
    GRS1758$-$258: flux in
    $\mathrm{keV}\,\mathrm{s}^{-1}\,\mathrm{cm}^{-2}$ in the
    3--10\,keV and 10--20\,keV bands, fitted to the spectra, spectral
    hardness (10--20~keV / 3--5~keV) calculated with fluxes in
    $\mathrm{keV}\,\mathrm{s}^{-1}\,\mathrm{cm}^{-2}$, photon index,
    temperature, and normalization of the disk component, total flux
    of the additional iron line in
    $10^{-3}\,\mathrm{ph}\,\mathrm{s}^{-1}\,\mathrm{cm}^{-2}$ and the
    reduced $\chi^2$. Soft states are highlighted for episodes
    reaching a photon index greater than 2.}
  \label{fig:soft}
\end{figure*}
To analyze the long term behavior of \grs, for each spectrum, the
source flux (i.e., with the Galactic Ridge background subtracted but
all other model components taken into account) was calculated
integrating over the best fit model in the respective energy ranges
for each observation. As apparent in Fig.~\ref{fig:soft}, a change in
flux is correlated with a change in the photon index: Once the flux
starts to decrease, the spectrum begins to soften. We classified as
soft state all episodes that reach photon indices greater than two.
The regions of interest were then defined to contain at least three
data points before and after the peak, to fully catch the rise and
decline. In some cases, this was not possible due to data gaps, which
we do not want to cross for lack of information, for instance after
the first soft state in 1999. Between 1997 and 2008, we found seven
dim soft states, which are highlighted in Fig.~\ref{fig:soft}. During
the 2001 soft state, the source almost turned off completely with a
remaining flux of only
${\sim}0.045\,\mathrm{keV}\,\mathrm{s}^{-1}\,\mathrm{cm}^{-2}$ in the
3--20\,keV band. The blackbody disk emission appears only during these
soft states where the low flux increases the uncertainties of the best
fit parameters. Looking at the reduced $\chi^2$ (see
Fig.~\ref{fig:soft}, bottom panel), we find the fits slightly
overdetermined, both for the fits with the fixed (Fig.~\ref{fig:soft})
and the free iron line width. A further reduction of free parameters,
however, is not possible: The Galactic ridge background is added as a
constant with no free parameters\footnote{Variable ridge emission is
  in principle possible due to changes in roll angle and possible
  transient background sources, however, spectral modeling in which we
  kept the spectral shape of the ridge constant but let its flux vary
  does not lead to appreciable changes in the results presented in
  this paper compared to models where the ridge emission was kept
  fixed.}. The iron line position is fixed, as is the absorption
towards \grs
\citep[$\nH = 1.5 \times
10^{22}\,\mathrm{cm}^{-2}$,][]{mereghetti:97}. The only free model
parameters are the powerlaw normalization and photon index, the iron
line flux, and the disk flux and temperature where a disk black body
is needed. Furthermore, we did not add a systematic uncertainty to the
data. We therefore conclude that we can not improve on the
overdetermination of our fits.

\subsection{Hardness intensity diagram}\label{sec:hid}

As shown, for example, by \citet{fender:04} or \citet{belloni:05}, it
is typical for black hole transients to trace a \textsf{q}-shaped
curve on their hardness intensity diagram (HID) during their
outbursts. \citet{pottschmidt:08} already found that \grs displays an
unusual behavior in this respect: while persistent binaries usually
occupy only a small area of the HID \citep[see also][]{wilms:07}, \grs
shows a mixture of transient and persistent behavior. It moves
anti-clockwise from the hard to the hard intermediate state, softens
and then dims to the soft state, and finally hardens along the lower
transitional branch back to the hard state. The 2001 extremely faint
soft state directly follows an observational gap. Therefore, the
transition from hard to soft state is not observed, leading to the
atypical shape in the HID \citep{pottschmidt:08}.

\begin{figure}
  \resizebox{\hsize}{!}{\includegraphics{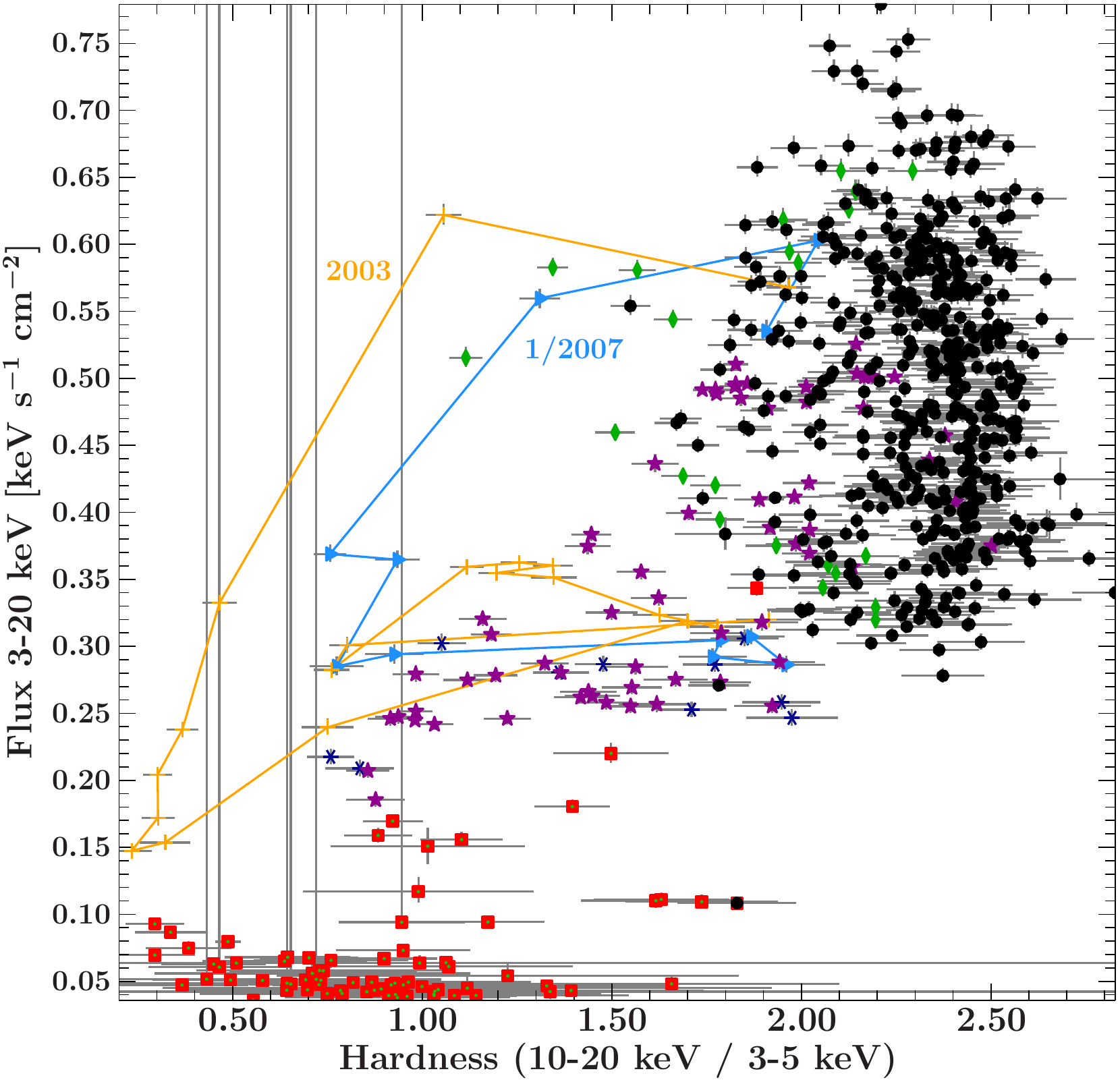}}
  \caption{Hardness intensity diagram (HID) from \xte monitoring
    observations of \grs from 1997 until 2007. The seven dim soft
    states are highlighted as in Fig.~\ref{fig:soft}. To show the
    ``\textsf{q}''-shaped track of \grs in the HID, the data points of
    two soft state passages (2003 and 1/2007) are connected.}
  \label{fig:hid}
\end{figure}

This behavior is confirmed here. Figure~\ref{fig:hid} shows the HID
for the whole \xte campaign. The soft states are highlighted using the
same color scheme as in Fig.~\ref{fig:soft}. It is obvious that \grs
does not follow the usual \textsf{q}-shaped track but rather starts
from the position of persistent sources on the upper right edge.
Although the HID shows a clear hysteresis for hard and soft (absorbed)
fluxes, there is no indication at all for a rise in the hard state
from quiescence. No full return to the hard branch could be observed
during the most extreme 2001 soft state: after the last soft state
data point (MJD 52235.2961), there are no observations for almost two
months. This mixed persistent/transient behavior of \grs was already
observed by \citet{smith:01}, similar behavior has also been seen in
1E\,1740.7$-$2942 \citep{delsanto:05} and GX\,339$-$4
\citep{delsanto:08}.

\section{Time series analysis}\label{sec:timing}

For the time series analysis we used the complete flux light curve of
\grs, that is the sum of the flux band light curves shown in
Fig.~\ref{fig:soft}: as opposed to the count rate, flux values are
independent of different detector responses to the respective spectral
shape of \grs. Due to many gaps and uneven spacing of the data points,
we had to use the algorithm for generalized periodograms after
\citet{lomb:76} and \citet{scargle:82}.

\subsection{Flux light curve and its periodogram}\label{sec:psdraw}

\begin{figure}
  \resizebox{\hsize}{!}{\includegraphics{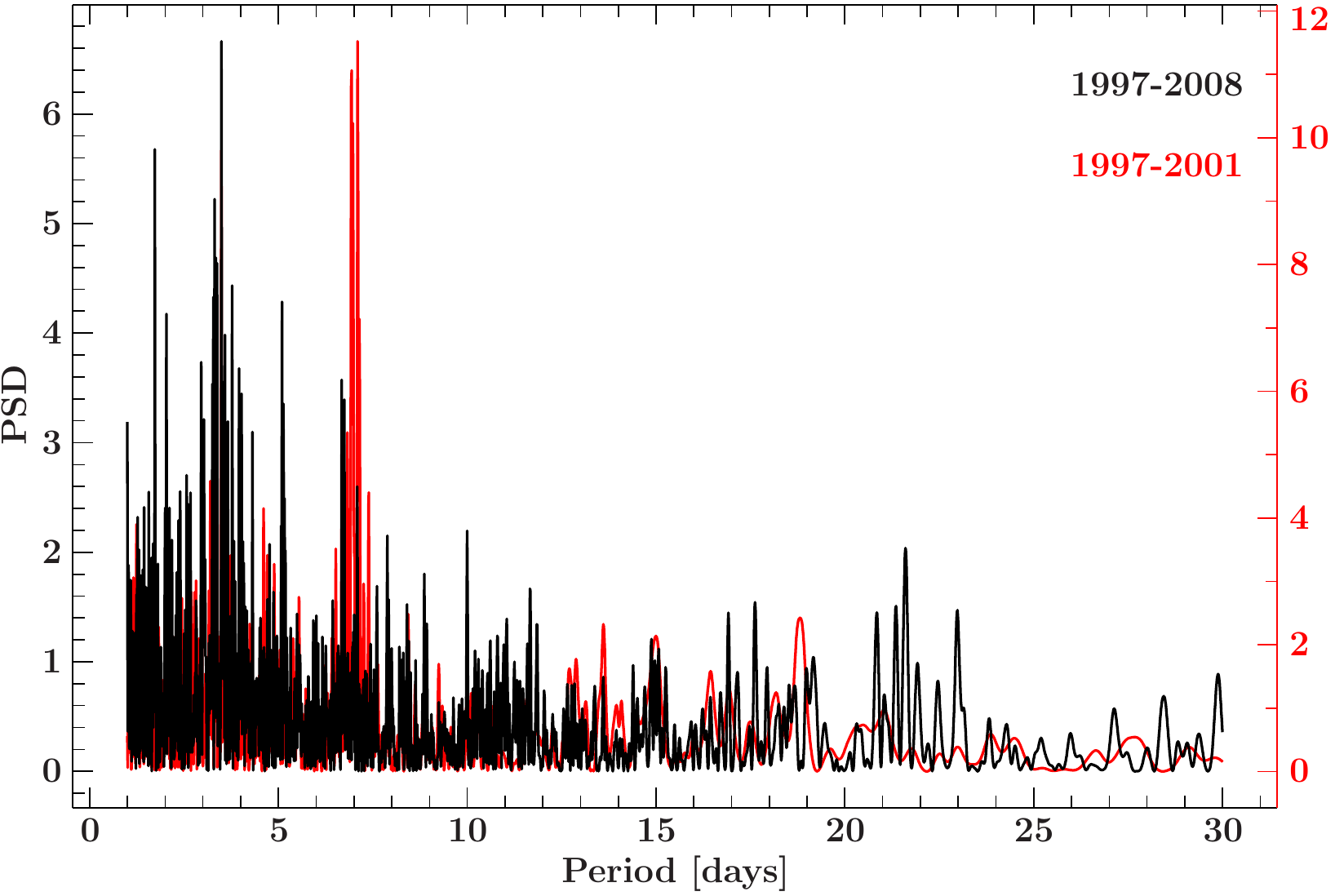}}
  \caption{Black: Power spectrum of the full and unfiltered light
    curve, and red: power spectrum of the unfiltered 1997--2001 data
    already used by \citet{smith:02}.}
  \label{fig:psd_unfiltered}
\end{figure}

We now turn to the search for (quasi-)periodicities in the long-term
light curve of the source, concentrating on the behavior of the
$18.45\pm0.10$\,d periodicity found by \citet{smith:02} in the
1997--2001 \xte-monitoring. Applying the Lomb-Scargle algorithm to the
whole unfiltered flux light curve led to a power spectrum without any
prominent peaks, and neither does the power spectrum calculated for
the 1997--2001 data contain significant peaks
(Fig.~\ref{fig:psd_unfiltered}). The large luminosity variations
between hard and soft spectral states can decrease the significance of
period measurements and therefore cause the lack of such peaks
\citep{smith:02}. In a next step we therefore perform a period search
on de-trended data under consideration of possible systematic effects.

\subsection{Detrended flux light curve and its periodogram: a drifting
  18\,day period}\label{sec:scargle}

Excluding all soft states from the time series analysis would lead to
major gaps in the light curve. In order to avoid any influence of the
very dim and soft states on our periodogram, we only used data points
with a photon index $\Gamma < 2$ (see Sect.\,\ref{sec:spectral}). This
step alone is not sufficient, however, to remove long term variations
and consequently no signal is seen in these power spectra. To avoid
lower significances of our measurements caused by large luminosity
variations and to be able to compare our results to those of
\citet{smith:02}, a high pass filter was then applied to the data by
subtracting a smoothed version of the light curve. The following
analysis uses this long-term trend and the high frequency residuals.

\begin{figure*}
  \centering
  \includegraphics[width=8.5cm]{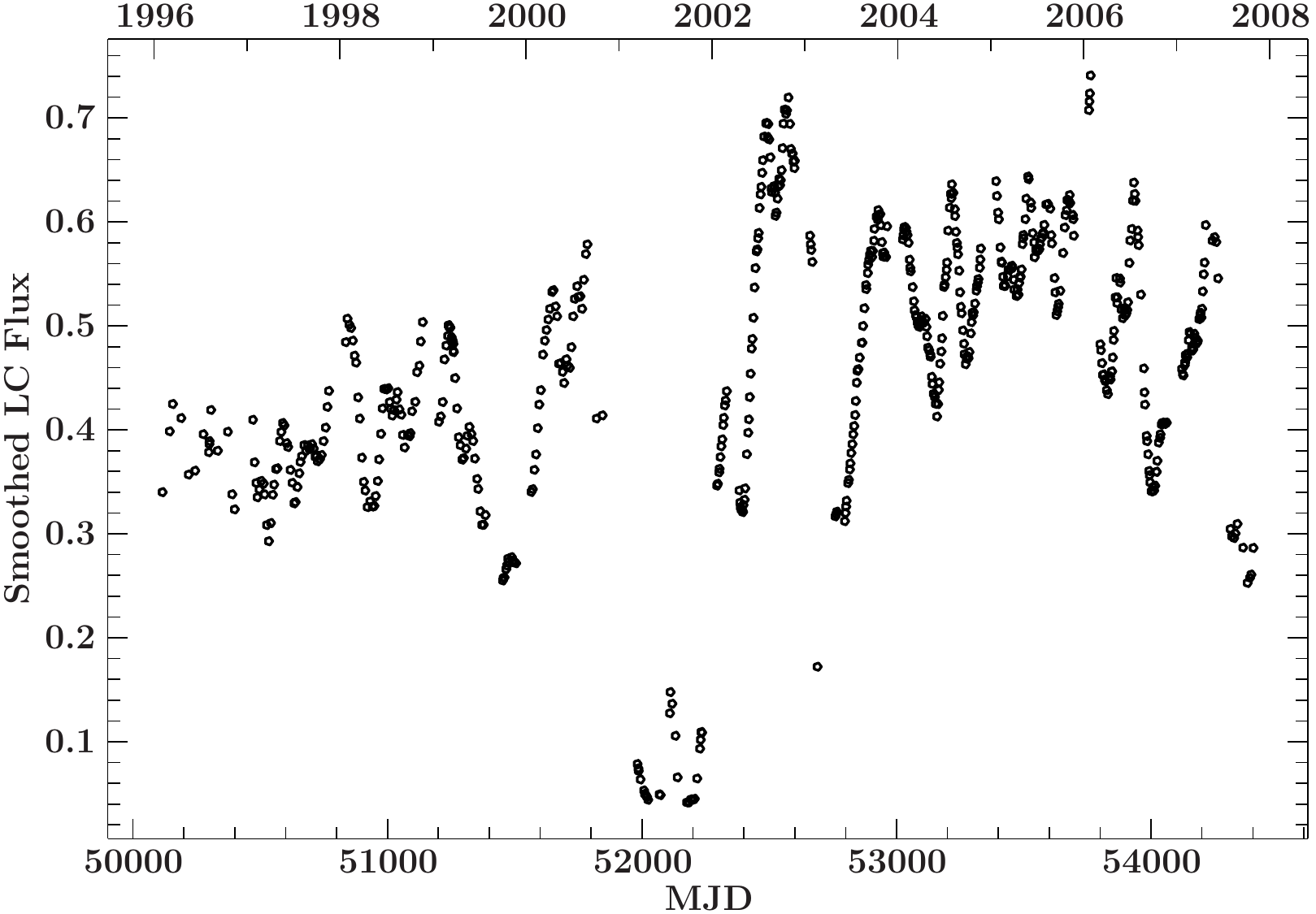}
  \includegraphics[width=8.5cm]{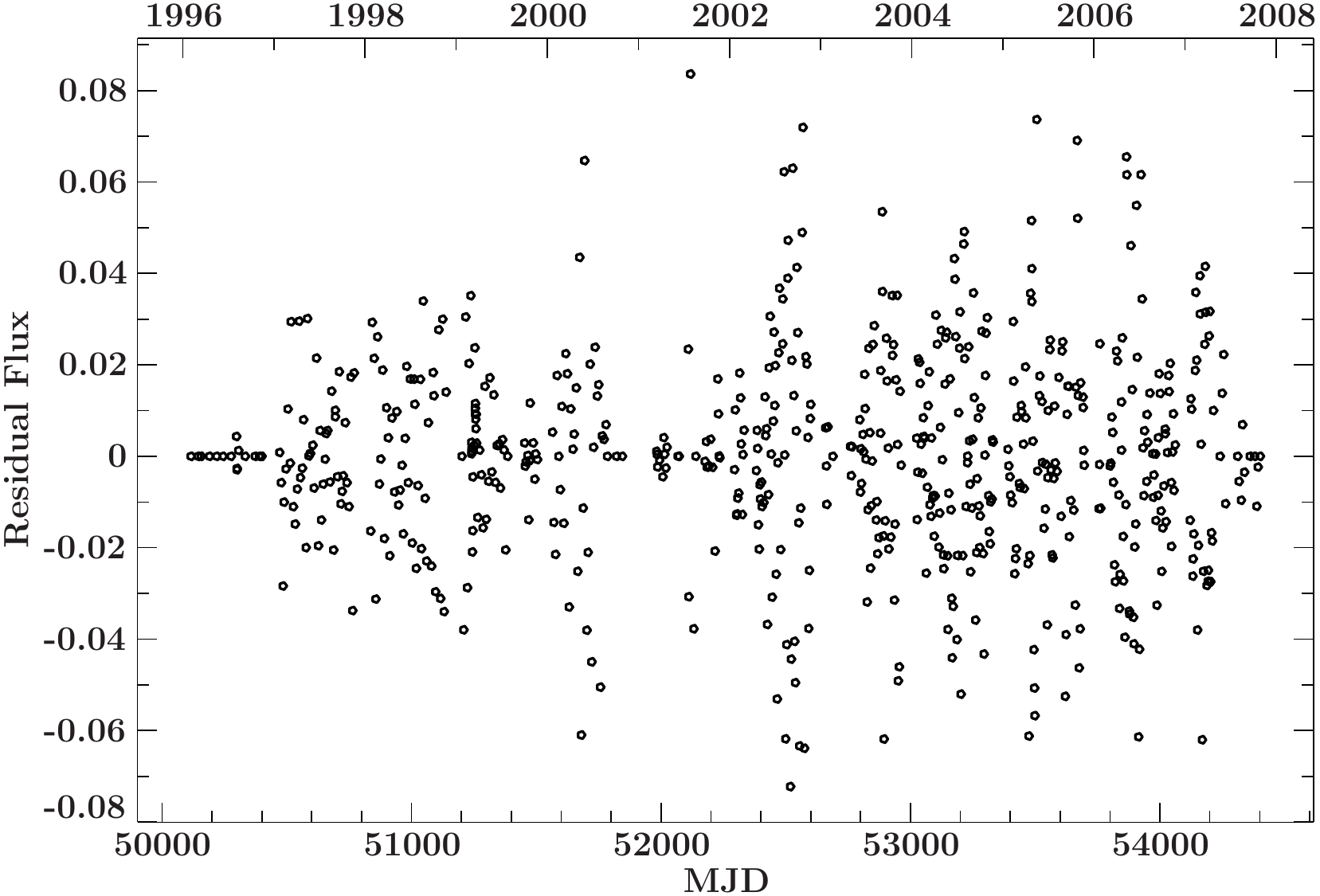}
  \caption{Long term trend (left) and residual flux (right) light
    curves in $\mathrm{keV}\,\mathrm{s}^{-1}\,\mathrm{cm}^{-2}$ after
    application of the high pass filter.}
  \label{fig:lc_res}
\end{figure*}

To generate the smoothed light curve, for each data point we fitted a
straight line to all data within the range of $n = 14$\,d before and
after. The subsequent analysis only used the high frequency residual,
that is the difference between the data point and the value of the
straight line to obtain a high pass-filtered light curve. This method
was already applied by \citet{smith:02}, who used a range of
$n = 10$\,d before and after the data point. We extended this range in
order to get better statistics for the smoothing fit, although we
emphasize that our results do not depend on the exact value of $n$
chosen \citep{hirsch:14b}. The smoothed long term trend and the
residual flux light curve are shown in Fig.~\ref{fig:lc_res}.

Using residual fluxes in the same time range as \citet{smith:02}, that
is 1997--2001, we are able to reproduce within the uncertainties the
peak they found at $18.45 \pm 0.10$\,d (Fig.~\ref{fig:lc_res}):
although we did not exclude the low energy flux where no modulations
are expected, we find a peak at $18.475 \pm 0.017$\,d.\footnote{While
  \citet{smith:02} calculate their uncertainty using the FWHM of the
  peak in the PSD, we used 1000 sets of the long term trend light
  curve plus Poisson-distributed random values for the respective
  light curve plus a sinusoid test signal at 16\,d. The standard
  deviation of the distribution of the PSD peaks of this test signal
  was then taken as a measure for our period uncertainty.}

\begin{figure*}
\centering
\includegraphics[width=17cm]{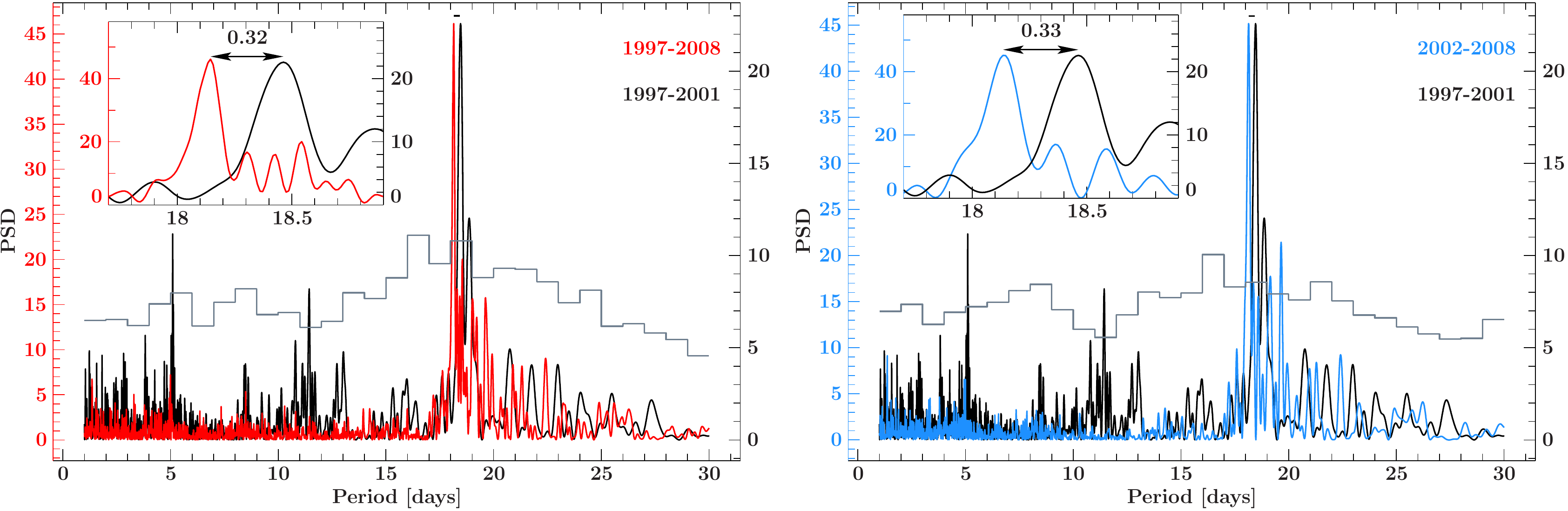}
\caption{Comparison of power spectra measured for the time interval
  1997--2008 (red, left) and 2002--2008 (blue, right) to the power
  spectrum measured in the 1997--2001 data (black), i.e., the time
  interval used by \citet{smith:02}. The peaks show a difference in
  period of 0.32\,d and 0.33\,d, respectively. Histograms show
  the peak PSD values for 10000 realizations of a detrended light
  curve with residuals with a Gaussian distribution in the indicated
  period interval (see text for further explanation).}
\label{fig:psd}
\end{figure*}

When using the whole 11\,years of data, however, this peak is shifted
in period by 0.32\,d (Fig.~\ref{fig:psd}, left). Analyzing the data
from 2002--2008, that is all data after the interval used by
\citet{smith:02} and therefore statistically independent from their
sample, this shift increases to 0.33\,d (Fig.~\ref{fig:psd}, right).
This difference is reminiscent of changing superorbital modulations
\citep[see, e.g.,][]{clarkson:03b,clarkson:03a}. The figure also shows
the maximum value of power spectra obtained when replacing the
residual with Gaussian noise with the same mean and standard
deviation. As discussed, for example, by \citet{benlloch:01}, these
lines represent the ``local significance'' that the observed
(quasi-)periodicity seen in the indicated 1\,d broad period intervals
is real. The Monte Carlo analysis automatically takes the trials
factor into account. The power spectra therefore show significant
peaks in the 18\,d period band.

\begin{figure*}
\sidecaption
\includegraphics[width=12cm]{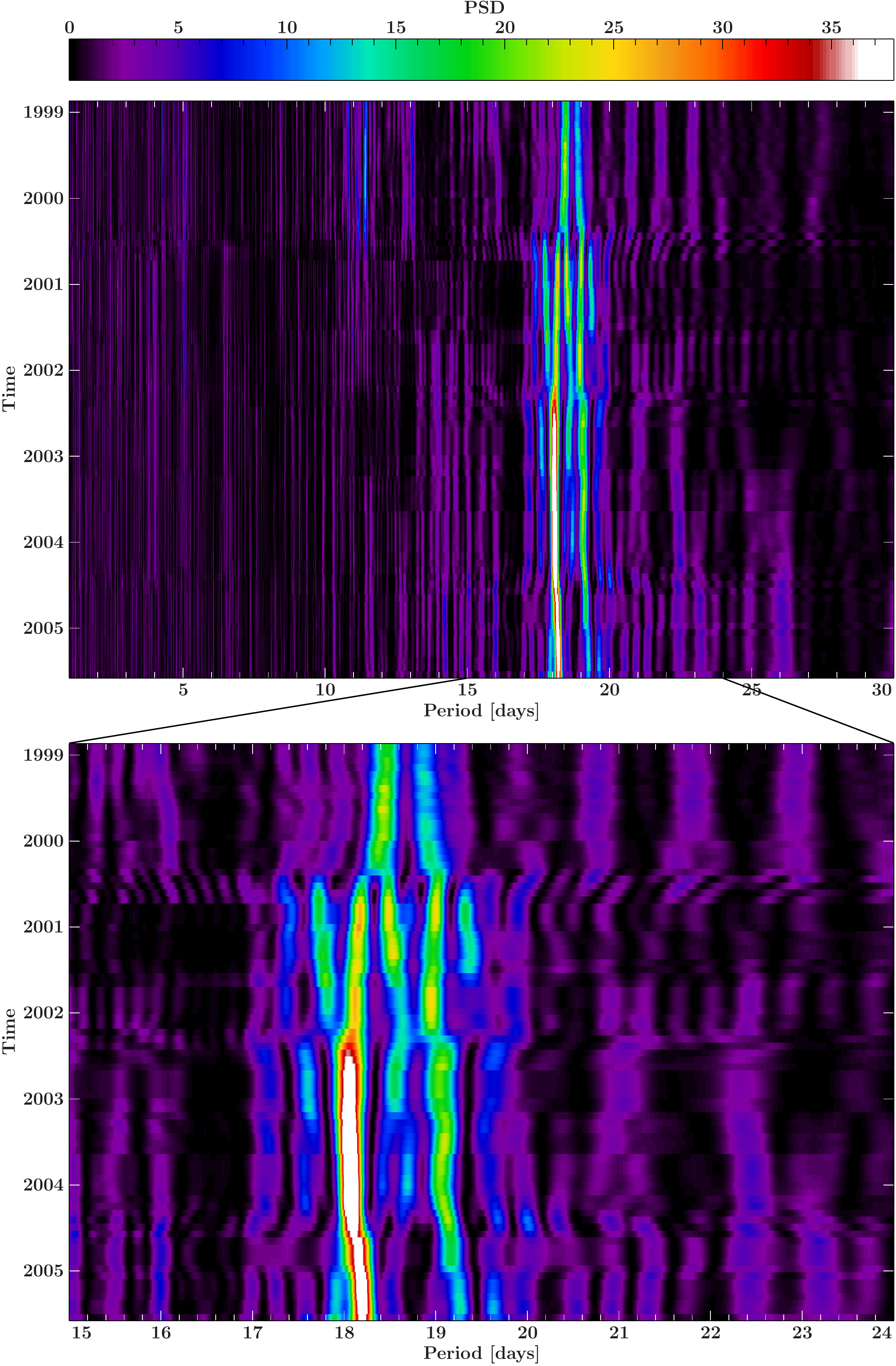}
\caption{Dynamical Lomb-Scargle periodogram for the entire 11\,year
  3--20~\keV model flux light curve of \grs. The periodogram is
  calculated in five year intervals (centered on the middle in the
  Time axis), stepped in intervals of 30\,d.}
\label{fig:dynpsd}
\end{figure*}

\begin{figure}
  \resizebox{\hsize}{!}{\includegraphics{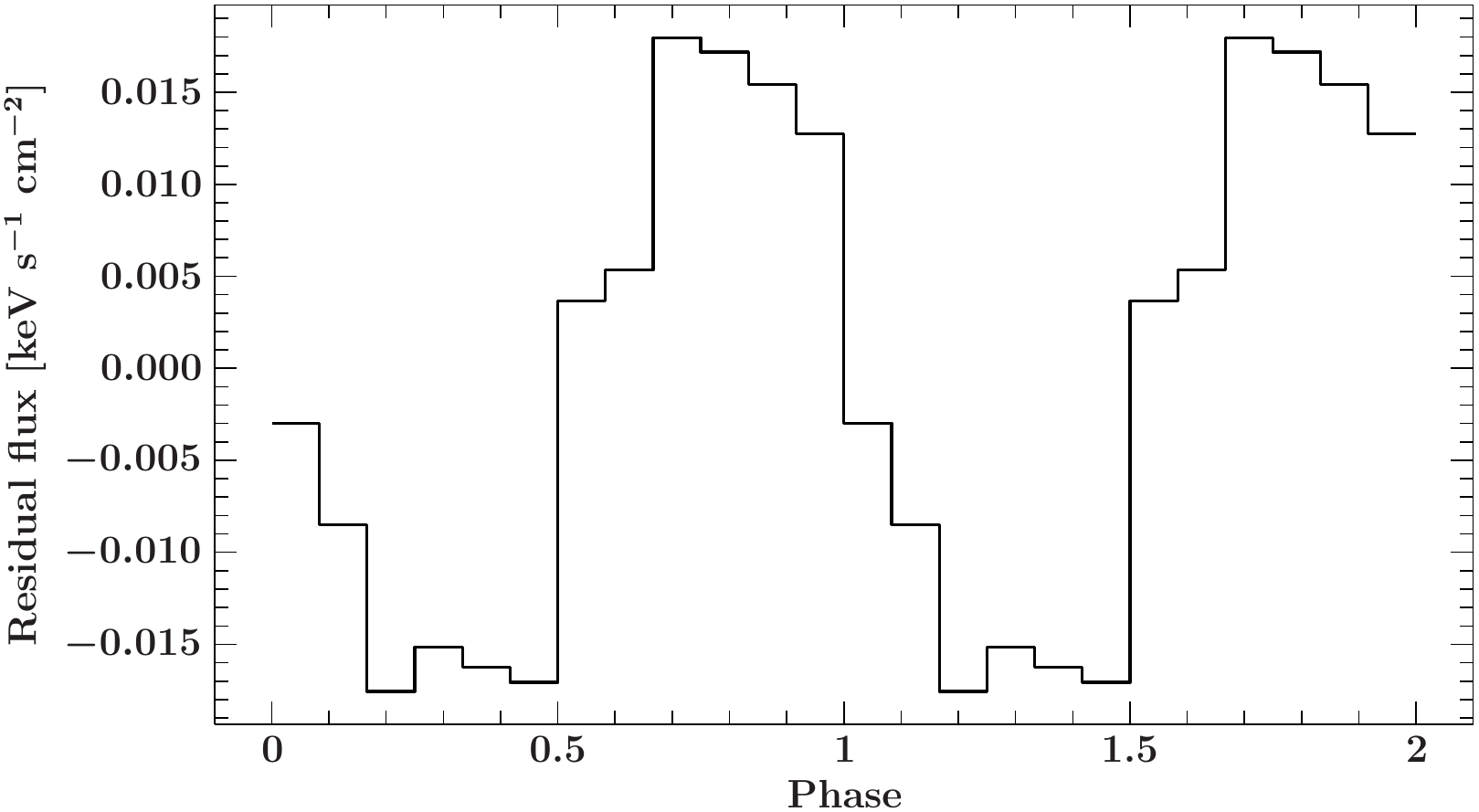}}
  \caption{Filtered 3--20~\keV model flux light curve for the stable
    part of the dynamical power spectrum (MJD 51769--53980), folded on
    a period of 18.09\,d.}
  \label{fig:foldedlc}
\end{figure}

To study the evolution of the quasi periodic signal, we calculated a
dynamic power spectrum \citep[Fig.~\ref{fig:dynpsd}, see
also][]{benlloch:01,wilms:01}: based on the 5\,years interval of data
originally used by \citet{smith:02}, slices of the same length of
5\,years were cut out of the light curve and analysed separately. Each
time the starting time of the slice was shifted by 30\,d, and each
resulting power spectrum is shown as a color-coded line in
Fig.~\ref{fig:dynpsd}. Note that the 83 individual slices are
overlapping and thus not statistically independent. As expected, the
first few lines of the dynamic periodogram show a peak at
$18.475 \pm 0.017$\,d. The maximum of this peak, however, is drifting
with time, first to lower values to a minimum of $18.043 \pm 0.005$\,d
in 2003. Towards the end of the light curve, the period of the main
peak is rising again. In addition, the period appears to fork into two
peaks between 2000 and 2001. Figure~\ref{fig:foldedlc} shows the
modulation of the filtered light curve for the time range where the
period in the dynamical power spectrum is stable (MJD 51769--53980).

Given the complexity of the data reduction, before we discuss the
scientific implications of the peak we first discuss the significance
of our measurements and explore how such a peak could be produced
artificially.

We start by estimating the significance of the period. In order to do
so we simulated 10000 light curves consisting of a white noise
component, that is Gaussian distributed random values with the same
standard deviation as that measured from the short term residuals
obtained after detrending the original lightcurve. This approach
therefore addresses both, the uncertainty of the individual flux
measurements due to statistical effects as well as any excess noise
that is due to intrinsic source variability. The residuals are
consistent with Gaussian noise, such that more complex modeling, for
example, using red noise residuals or applying a bootstrapping
approach, is not necessary. For each simulated light curve, a dynamic
power spectrum with 83 overlapping light curve slices was calculated
in the same way as in Sect.~\ref{sec:scargle}. For each light curve
slice, we then find the fraction of simulated white noise light curves
that do not have their highest PSD peak in the range of the respective
detected \grs period and its uncertainties. This fraction is a measure
of the significance of the putative period of \grs. We find
significances for the drifting peak varying between 98.15\% and
99.98\%.

Although our high pass filter works well for detecting periodic
signals in the data, it is not ideal. For example, the long term trend
light curve, which is subtracted by the filter, still contains part of
periodic signals in the range of 14\,d to 25\,d. As the period found
by \citet{smith:02} falls in this range, a closer look into the origin
of this effect of the filter on the trend light curve is required.

To evaluate the range in which remnants of short-term periodic signals
can be found in the long-term trend curve, we build an artificial
light curve sampled at the same dates as those of the original light
curve, containing Gaussian noise and a primary sinusoidal signal with
a period in the range of 14\,d to 25\,d. This light curve is then
filtered to determine the long term trend. To test whether this trend
still contains parts of the primary signal, we added an additional
test signal at a different period, for example 12\,d, filtered again
and applied the Lomb-Scargle technique as in Sect.~\ref{sec:scargle}.
The resulting periodogram shows a main peak at the period of the test
signal and a second peak at the period of the primary sinusoidal
period. Since the analysis is based on a noise light curve, the
residual signal is not connected to the flux values in our \grs light
curve. Neither changing the period of the test signal nor changing the
length of the light curve section we use has any influence on this
effect. Randomly selecting two thirds of the data points in the light
curve lessens the effect, and at the same time also reduces the power
seen in the test signal peak.

The distribution of time intervals between individual observations of
\grs, that is the time intervals between the data points in the \grs
light curve, does not show an excess for intervals between 14 and
25\,d. We can therefore exclude that the periodicity found in the
long-term trend light curve is the result of the sampling of our \grs
light curve. However, the range, in which the residual signal appears
in the trend light curve, shifts according to the filter range $n$ and
is always located between $n$ and $2n$. Thus we conclude that this
residual signal is left because the filter is not an ideal high pass
filter. But as we find a significant peak in the periodogram although
the filter removes part of the signal together with the long term
trend, the filtering approach effect does not impair the main results
of our analysis.

We also tested whether the drifting peak in the \grs dynamic power
spectrum could be caused by the filtering process. In order to do so,
we consider a pure white noise light curve with the original sampling,
apply the high pass filter, and then calculate a dynamic power
spectrum of the short term residuals analogous to the \grs dynamic
power spectrum shown in Fig.~\ref{fig:dynpsd}. As expected, none of
the light curve slices, that is lines in the dynamic power spectrum,
shows a prominent peak at periods between 1\,d and 30\,d in its
periodogram. Thus, the period is not created by the filtering process.

Finally, in order to determine whether the shifting peak is real, we
also tested whether the peak could be caused by the variability of the
source being red noise. Red noise, used here to mean a stochastic
process with a $f^{-\alpha}$ power spectrum, is notorious to exhibit
quasi periodicities in its lightcurve when studying short light curve
segments. We therefore used the algorithm of \citet{timmer:95} to
generate a red noise lightcurve with 100000 data points (in order to
avoid windowing effects caused by the simulation approach) with the
same overall statistical properties as that of \grs. We then selected
a segment of that light curve to generate a red noise only lightcurve.
This light curve was then analyzed in the same way as the real data,
again using a high pass filter before applying the Lomb Scargle
algorithm. The resulting dynamic power spectrum does not show any peak
at all in the period range we are interested in. We therefore can
exclude a red noise origin for the drifting peak.

\subsection{Comparison with \citet{levine:11}}\label{sec:psdasm}

\citet{smith:02} were not the only ones reporting a periodicity for
\grs: Using ASM data in the 3--5\,keV band and a different filtering
approach, \citet{levine:11} found a signal at a frequency of
$0.0527\,\mathrm{cycles}\,\mathrm{day}^{-1}$, corresponding to a
period of 18.97\,d. As this value is slightly different from what we
obtain using the method of \citet{smith:02}, we tried to reproduce
their result, implementing the analysis method as described in
\citet{levine:11}: We use the ASM light curve rebinned to a 2\,d
resolution and weighted according to the description in
\citet{levine:11}, smooth it with a Gaussian kernel function with a
full width at half maximum of 500\,d, and then calculate the power
spectrum via the classical Fourier transform. This power spectrum is
then whitened to account for background power. Detailed descriptions
of all these steps can be found in the appendix of \citet{levine:11}.
The reason for using a different approach than for the data described
in Sect.~\ref{sec:scargle} is twofold: First, it allows us to see
whether the results of \citet{levine:11} hold also for the longer time
interval considered here. Secondly, using a different methodology on
the ASM data set avoids introducing the same potential systematic
errors in the analysis and thus allows an independent confirmation of
the results of the PCA analysis of Sect.~\ref{sec:scargle}.

\begin{figure}
  \resizebox{\hsize}{!}{\includegraphics{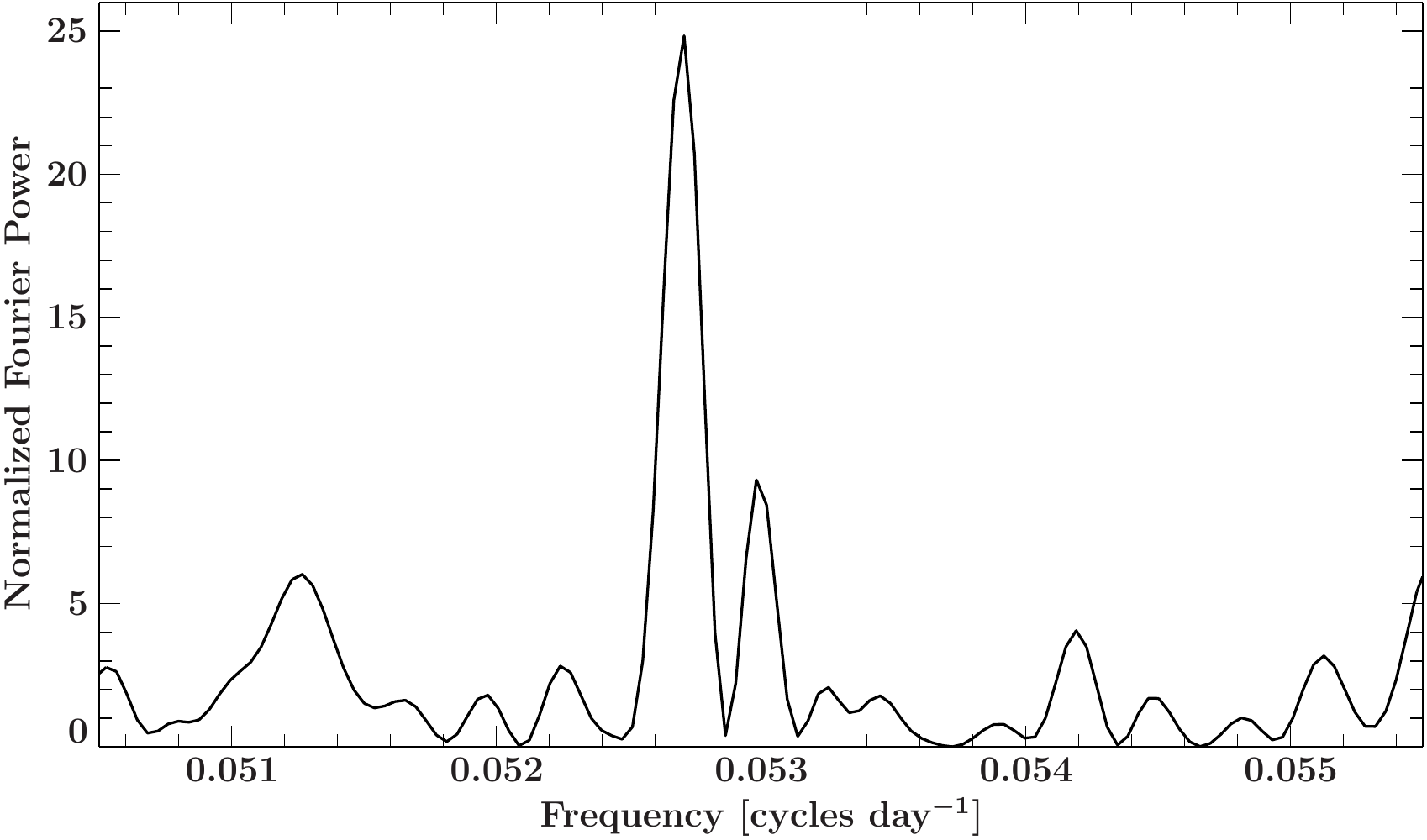}}
  \caption{Power spectrum of the 3--5\,keV ASM light curve, using the
    method of \citet{levine:11}. We are able to reproduce their
    Fig.~9 (bottom).}
  \label{fig:levinefig9}
\end{figure}

\begin{figure}
  \resizebox{\hsize}{!}{\includegraphics{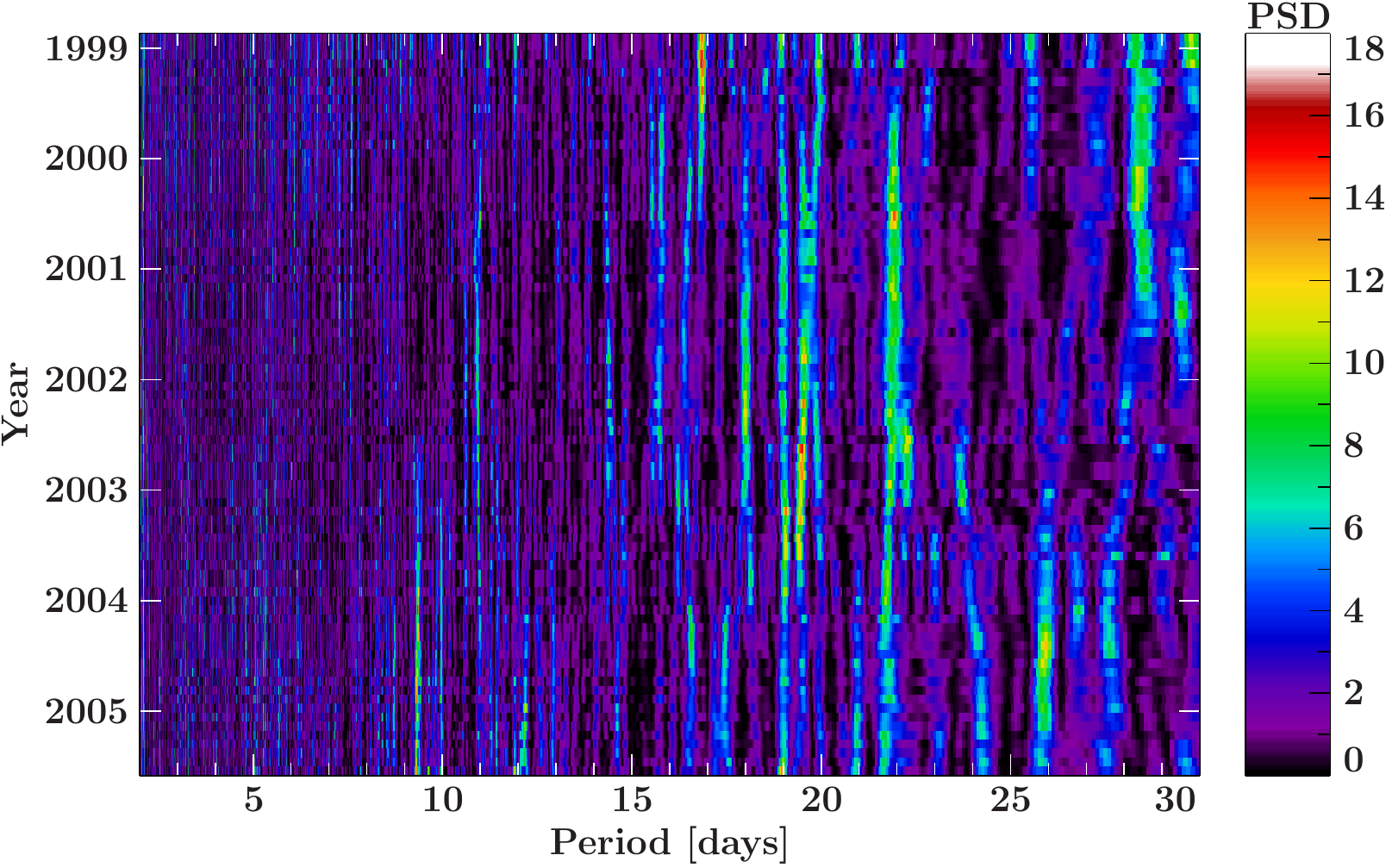}}
  \caption{Dynamic power spectrum of the 3--5\,keV ASM light curve,
    using the method of \citet{levine:11}. A drifting peak is visible
    although much less prominent due to the poor signal to noise ratio
    of the ASM data.}
  \label{fig:asm_dynamic}
\end{figure}

Using this method, we are able to roughly reproduce their power
spectrum of \grs \citep[][their Fig.~9, bottom]{levine:11}. There is a
deviation in the total power in the power spectrum which is due to
differences in the normalization of the Fourier transformation
routines (Fig.~\ref{fig:levinefig9}). We then calculated a dynamic
power spectrum (Fig.~\ref{fig:asm_dynamic}) for the same light curve
slices as above, using the ASM 3--5\,keV light curve binned on a 2\,d
grid and the analysis method of \citet{levine:11}. As the ASM data are
noisier than the PCA flux light curve (Fig.~\ref{fig:dynpsd}), there
is no prominent peak as seen in the PCA data. However, there is a
feature around 19\,d as reported by \citep{levine:11}. This feature
blends with a peak that starts at a period of nearly 20\,d and then
drifts first toward smaller periods with a minimum in 2003. As with
the PCA data and analysis after \citet{smith:02}, a side peak showing
a similar behavior at longer periods is also present. However, in the
ASM data the side peak is at a distance of $\sim$2\,d from the main
peak, while in the PCA data the distance is only $\sim$1\,d. The
region around mid-2000, where the main peak of the PCA data splits up
into two peaks, is also interesting in the ASM data: the main peak
drifts towards a minor peak and broadens as the two periods are close
to each other.

On the whole, however, apart from a systematic discrepancy in periods
of about 1\,d, the analysis of the ASM data using this different
method confirms our first result of a drifting periodicity for \grs:
It is seen in different datasets, analyzed with different methods.

\begin{figure}
  \resizebox{\hsize}{!}{\includegraphics{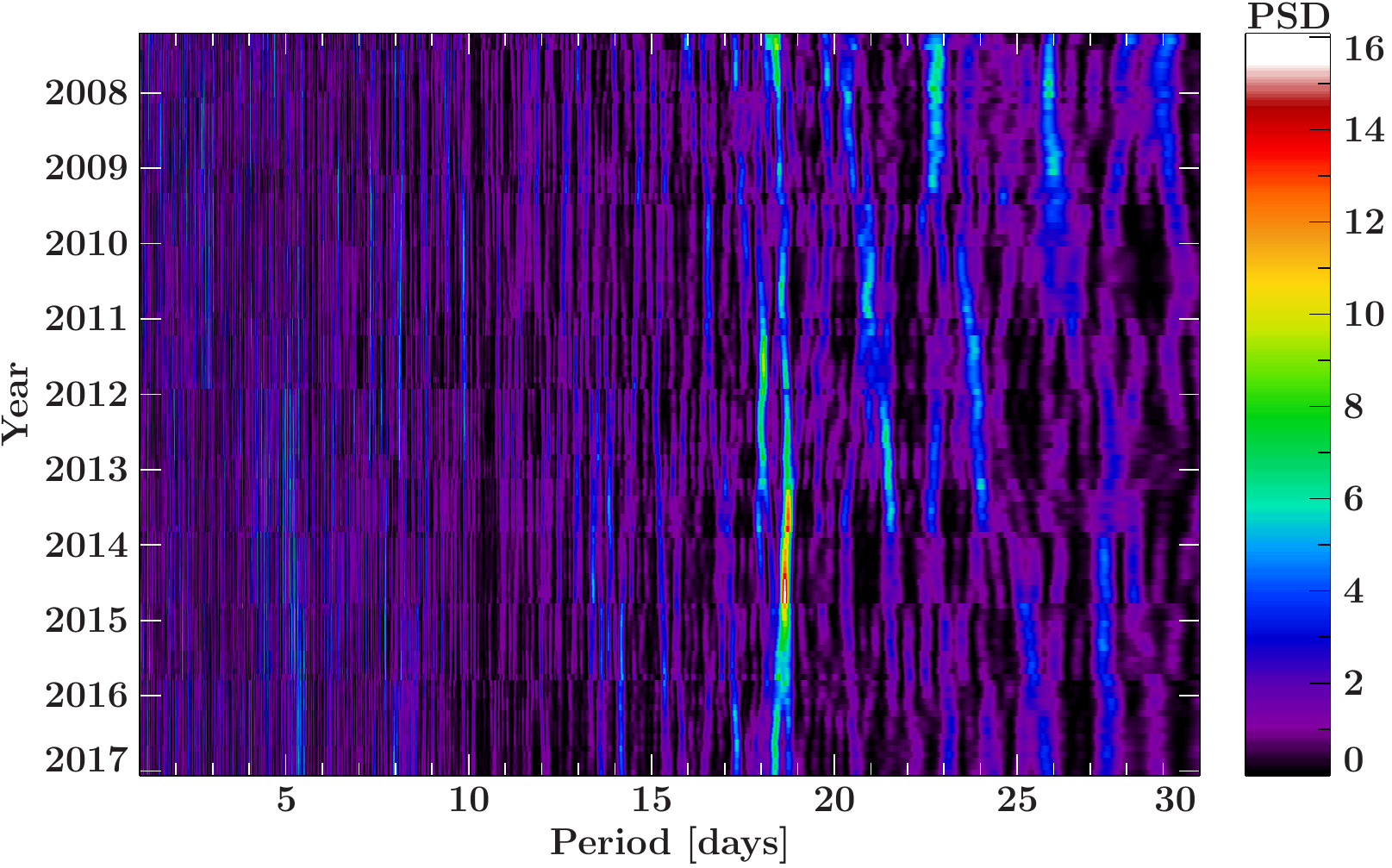}}
  \caption{Dynamic power spectrum of the 15--50\,keV \textsl{Swift}
    BAT light curve, using the method of \citet{smith:02}. The high
    energy light curve also shows a drifting peak. A change in period
    is obvious in 2016, near the most recent soft state.}
  \label{fig:dynpsd_bat}
\end{figure}

For another cross-check we analyzed the \textsl{Swift} BAT light curve
taken from the Transient Monitor \citep{krimm:13} following the
algorithm of \citet{smith:02}, using the same parameters for the high
pass filter as in the analysis of the \xte data. The resulting dynamic
power spectrum is shown in Fig.~\ref{fig:dynpsd_bat}. The energy band
of 15--50\,keV also shows a drifting periodicity between 18\,d and
19\,d. The signal is not as strong as in the PCA data, but the
behavior is clearly the same in high as in low energies. Note that
around the most recent, very dim 2016 soft state \citep[][Hirsch et
al., in prep.]{pottschmidt:16,hirsch:16} a decrease in period is
visible similar to the decrease in the PCA dynamic power spectrum
around the very dim 2001 soft state. This additional observation of
the drifting periodicity with another satellite than \xte, at another
time, in another energy range, further increases the confidence that
we are seeing a source-intrinsic signal.

\section{Discussion}\label{sec:summary}

\subsection{Spectral evolution and state transitions}
Analyzing data of the eleven years long \xte-PCA monitoring of \grs we
find that the spectrum in the 3--20\,keV range can always be described
by an absorbed powerlaw (photon indices varying between 1.5 and 3)
with a neutral Fe K$\alpha$ line (which might be due to variations in
the Galactic ridge emission) and, during the soft states with a photon
index softer than 2, a disk-blackbody component. The Galactic ridge
background emission was always accounted for. \grs entered a dim soft
state for seven times between 1997 and 2008. During these soft states,
the track of \grs in the hardness intensity diagram is similar to the
q-shaped one of transient sources, however there is no rise from
quiescence into the hard state. The high energy HID of \grs is
comparable to that of \cyg \citep{obst:11b}.

The detection of an accretion disk in the soft state and its
non-detection in the hard state are consistent with the
\textsl{XMM-Newton} observations discussed by \citet{soria:11}, who
also find a higher total X-ray luminosity in the 2001 \textsl{XMM}
soft state data than in the 2008/2009 \textsl{Swift} XRT hard state
observations. They conclude that luminosity cannot be the only driving
force for a state transition in the \grs system.

There are several models that try to explain the evolution of X-ray
binaries in the HID. One of the first attempts to explain the
hysteresis is the picture of two accretion flows set up and applied to
\grs by \citet{smith:02a}, and similar to models discussed by
\citet{meyer:00a} and \citet{meyer-hofmeister:09}: Based on
\citet{chakrabarti:95}, they suggest a Keplerian accretion disk in
combination with a hot, sub-Keplerian halo accreting proportional
amounts of matter. A boost in accretion rate leads to the halo
brightening at once (almost free fall timescale), while the
brightening of the inner regions of the disk is delayed by the
inspiral of matter (viscous timescale). Then the additional soft
photons are cooling the halo and the spectrum softens. Similarly, a
sudden drop in accretion rate would first affect the halo, its
Comptonizing component declining rapidly. The Keplerian disk reacts
only on the viscous timescale, causing the soft component to decay
slower \citep{smith:02a,smith:07}, which is the ``dynamical soft
state'' observed in \grs and other black hole binaries, for instance,
the transient source GX\,339$-$4 \citep{debnath:15,nagarkoti:16} and
many other transients \citep{gierlinski:06} as well as the persistent
``twin source'' of \grs, \einsE \citep{smith:02a}. \citet{soria:11}
refined this idea and suggest a magnetically powered coronal outflow
as the source of the hard radiation. The accretion flow could then
switch between the hard corona and the soft accretion disk because of
changes in the poloidal magnetic field.

The interpretation of two accretion flows fits not only \grs but also
other Galactic black hole binaries, With small changes, this picture
can also be applied to high-mass X-ray binaries such as \cyg and
LMC~X-3 \citep{smith:02a,smith:07}: Here, the mass input is no longer
distributed proportionally between disk and halo. State changes are
induced by the accretion flow switching between favouring the disk
while starving the halo and favouring the halo while starving the
disk. Thus, the bolometric luminosity should remain almost unchanged
during state transitions in these systems. Recently, \citet{ghosh:18}
have found further evidence for an advective and a Keplerian flow
analyzing time lags in the spectral slope for different HMXBs and
LMXBs. They use a more extensive dataset and a completely different
method than \citet{smith:02}, yet their results lead to the same
conclusion.

We note, however, that other models are equally successful at
describing state transitions, such as the hybrid model of an outer
standard accretion disk and an inner, magnetized jet emitting disk of
\citet[][see also \citealt{ferreira:06},
\citealt{marcel:18a,marcel:18b}]{petrucci:08}, the explanation of the
hysteretic cycle in black hole state transitions as a magnetic field
effect put forward by \citet{begelman:14}, or the explanation of state
transitions through severe disruptions of the accretion flow by
\citet{nixon:14}. Common to all of these analyses and also to the
large number of other discussions in the literature is that they are
biased by the few bright and well sampled black hole outbursts such as
those from GX~339$-$4 or XTE J1550$-$564, and that the more complex
hysteretic behavior seen here or in other persistent sources such as
LMC~X-3 or Cyg~X-1 is typically not explained. We hope that the data
presented here will stimulate further theoretical discussions that
address the difference of \grs and other black hole binaries.

\subsection{Timing behavior and long-term evolution}

Analyzing the model flux light curve spanning 11\,years of
observations for timing analysis, we are unable to detect any orbital
modulation in the dataset. However, after detrending the data we find
that the dynamic power spectrum exhibits a peak which drifts at
periods between $18.475 \pm 0.017$ and $18.043 \pm 0.005$\,d and has a
significance between 98.15\% and 99.98\%. This drifting behavior was
confirmed using another data set and analysis method, however with a
systematic deviation in periods of about 2\,d.

Long-term periodicities in accreting systems are generally associated
with periodic phenomena in the outer parts of the accretion disk,
which are due to a combination of radiation pressure and orbital
effects. The most prominent of such radiation driven periodicities are
superhumps in cataclysmic variables (CVs). Such warps were first
observed during superoutbursts of SU UMa systems \citep[see][for a
review and, e.g., \citealt{armstrong:13a} for observations of
superhumps in several CVs]{warner:03}, where the superhumps are seen
as periodic optical modulations caused by a 3:1 orbital resonance
within the accretion disk, which causes the disk to be eccentric and
to slowly precess. Here, irradiation of the accretion disk by the
central source or inner part of the accretion disk results in a net
torque on the disk which leads to a precessing, warped disk
\citep{petterson:77, iping:90, pringle:96, wijers:99, maloney:96}. The
luminosity modulation is then caused by periodic variations of the
efficiency of dissipative processes in the accretion disk
\citep{whitehurst:88, whitehurst:91,lubow:91a,lubow:91b}.

\citet{masetti:96}, \citet{haswell:01}, and \citet{charles:02} review
observations of superhumps in soft X-ray transients and low mass X-ray
binaries. \citet{masetti:96} suggest an alternative origin for these
modulations: with an elliptical disk shape, the accretion flow impacts
the outer disk at varying distances from the central object and thus
at different gravitational potentials, leading to modulations in the
released energy. Other possible mechanisms include a variation in the
uncovered area in the direction of the observer or varying absorption
by a disk warp. Based on this idea, \citet{clarkson:03a} suggest a
similar mechanism to explain the superorbital period of the high mass
X-ray binary SMC~X-1 \citep{wojdowski:98}, which consists of a neutron
star and the B0 I optical companion Sk\,160 \citep{reynolds:93}.
\citet{clarkson:03a} found this period to be varying between 40\,d and
60\,d. These authors performed an analysis similar to ours, and also
their dynamic power spectrum looks similar. They suggest the
modulation being due to a bright spot at the intersection of accretion
flow and accretion disk. This mechanism can support variations in the
superorbital period \citep{clarkson:03a}. In a follow-up paper,
\citet{clarkson:03b} present the analysis of a sample of other sources
showing superorbital periods and put up a scheme, showing the
evolution of disk warping with respect to the binary radius. With
regard to the predictions of \citet{ogilvie:01},
\citeauthor{clarkson:03b} propose that warping due to irradiation of
the disk is impossible for very close binaries. With increasing
separation of the binary components one stable warp mode as seen, for
example, in Her~X-1 or LMC~X-4. Above the boundary region, several
strong periodicities interact, as seen, for example, in Cyg~X-2. In
the border region itself, stable warping is not possible, as is seen
in SMC~X-1, which shows sharp variations in the superorbital cycle
length \citep{trowbridge:07}. The underlying model, however, has to be
more complex than described by \citet{clarkson:03a}, who expect a
precessing warp and long periodicity for a source near this boundary,
which is clearly inconsistent with the behavior reported by
\citet{trowbridge:07}.

The result of \citet{clarkson:03b,clarkson:03a} raises the question
whether a similar mechanism is also applicable to \grs. Both low mass
X-ray binaries Her~X-1 and LMC~X-4 accrete via Roche lobe overflow,
and the high mass X-ray binary SMC~X-1, too, is best characterized by
Roche lobe overflow properties \citep[e.g.,][]{li:97,icdem:11}. Given
that the companion of \grs is probably an A-type star
\citep{marti:16}, the system would probably be somewhere in between
Her~X-1 or LMC~X-4 and SMC~X-1, such that the observed behavior is not
fully unlikely.

Beyond that, superorbital periods have also been observed in
wind-accreting high mass X-ray binaries \citep[e.g.,][]{corbet:13}.
For such systems, different mechanisms have to be considered.
\citet{koenigsberger:06} suggest oscillations in the companion star
driven by tidal interactions to be the source for the superorbital
period, while \citet{bozzo:17} propose corotating interaction regions
in the stellar wind to be responsible for the observed modulations.

There are many mechanisms that lead to an observable variability in
the light curve for different kinds of binary systems, and many
aspects that can influence the formation of warps in an accretion disk
for a system like \grs, so that we cannot finally conclude this
special mechanism of a warped disk to be the origin of the variable
periodicity in \grs.

\section{Conclusions}
Overall, the \xte monitoring data show that \grs fits into the general
picture of X-ray binaries with a few source-characteristic
features. The occasional very dim soft states as well as the striking
timing behavior put a challenge to the current available physical
models.
 
With the data currently available, it is neither possible to decide
between the different models for the state transition in \grs nor
between the mechanisms that lead to an observable variability in the
light curve of a binary system. Further steps in theory and simulation
have to be made to answer the open questions such as
\begin{itemize}
\item What physical model is behind the state transitions in \grs and
  other black hole binaries?
\item Does the same model apply to low mass and high mass X-ray
  binaries, or do we need different mechanisms?
\item Is there a model that can accomodate not only black hole
  binaries but also neutron star low-mass binaries, which display a
  similar behavior \citep{maccarone:03, munoz-darias:14}?
\item What factors influence the formation of warps in accretion disks
  in \grs and other sources and how do they influence it?
\item Is it possible to explain both effects in one comprehensive picture?
\end{itemize}

Especially for the last two items, further observations of systems
that are displaying superorbital periods are needed to have a
statistical relevant sample of different behaviors and to be able to
fit in the scheme sources as \grs where we do not know much about the
distance, the companion or the binary separation and orbit.

\begin{acknowledgements}
  This research has made use of ISIS functions provided by ECAP/Remeis
  observatory and MIT (http://www.sternwarte.uni-erlangen.de/isis/).
  We thank John E. Davis for the development of the SLxfig module,
  which was used to create all figures in the paper. This research was
  partially funded by the Bundesministerium f\"ur Wirtschaft und
  Technologie under Deutsches Zentrum f\"ur Luft- und Raumfahrt grant
  50\,OR\,1113. This work has been partially funded by the European
  Commission through grant ITN 215212 ``Black Hole Universe''. It was
  partially completed by LLNL under the auspices of the US DOE under
  Contract DE-AC52-07NA27344. We acknowledge the support by the DFG
  Cluster of Excellence ``Origin and Structure of the Universe'' and
  are grateful for the support by MCB through the Computational Center
  for Particle and Astrophysics (C2PAP). Support for this work was
  provided by NASA through the Smithsonian Astrophysical Observatory
  (SAO) contract SV3-73016 to MIT for Support of the Chandra X-Ray
  Center (CXC) and Science Instruments. CXC is operated by SAO for and
  on behalf of NASA under contract NAS8-03060. F.K.\ acknowledges
  funding from the WARP program of The Netherlands Organisation for
  Scientific Research (NWO) under grant agreement No~648.003.002, and
  was supported as an Eberly Research Fellow by the Eberly College of
  Science at the Pennsylvania State University. V.G.\ is supported
  through the Margarete von Wrangell fellowship by the ESF and the
  Ministry of Science, Research and the Arts Baden-W\"urttemberg.

\end{acknowledgements}

\end{document}